\documentclass{article}

\usepackage[margin=1in]{geometry}
\usepackage[affil-it]{authblk}
\usepackage{colortbl}
\usepackage{amsmath}
\usepackage{amssymb}
\usepackage{multirow}
\usepackage{graphicx}
\usepackage{xurl}
\usepackage{textcomp}
\usepackage{gensymb}
\usepackage{subcaption}

\definecolor{Gray}{gray}{0.9}

\title{\hrule height 1.8pt \vspace{0.3cm} \textbf{Decoupled coordinates for machine learning-based molecular fragment linking}\\\vspace{0.3cm}\hrule height 1.8pt\vspace{0.5cm}\large A Preprint\vspace{0.5cm}}

\author[1]{\textbf{Markus Fleck}\thanks{email: m.fleck@celeristx.com, corresponding author}}
\author[1]{\textbf{Noah Weber}\thanks{email: n.weber@celeristx.com}}
\author[1]{\textbf{Christopher Trummer}\thanks{email: c.trummer@celeristx.com}}
\affil[1]{Celeris Therapeutics GmbH\\
	Salzamtsgasse 7\\
	Graz, 8010}

\begin{document}
\maketitle
\begin{abstract}
Recent developments in machine-learning based molecular fragment linking have demonstrated the importance of informing the generation process with structural information specifying the relative orientation of the fragments to be linked. However, such structural information has not yet been provided in the form of a complete relative coordinate system. Mathematical details for a decoupled set of bond lengths, bond angles and torsion angles are elaborated and the coordinate system is demonstrated to be complete. Significant impact on the quality of the generated linkers is demonstrated numerically. The amount of reliable information within the different types of degrees of freedom is investigated. Ablation studies and an information-theoretical analysis are performed. The presented benefits suggest the application of a complete and decoupled relative coordinate system as a standard good practice in linker design.
\end{abstract}

\textbf{\emph{Keywords}} Fragment Linking $\cdot$ Deep Generative Models $\cdot$ Drug Design

\section{Introduction}
Computational drug design remains a challenging problem, foremost due to the vast size of the drug-like chemical space \cite{chemical_space_size, Reymond2010, Li2020, Mullard2017}. A subset of the discipline of molecular generation is constituted by the task of fragment linking. Given a pair of structures, the goal of the design process is their connection via an appropriate linker. Chemical properties like toxicity and solubility are of general concern in order for a compound to turn out administrable as a drug. Furthermore, in order to maintain high activity and specificity, it is of major importance that the designed linker maintains the binding mode of the underlying fragments \cite{fragment_linking1,fragment_linking2}. Therefore, an effective linker design process should incorporate structural information in the form of the relative orientation between the molecules to be joined. Recently, DeLinker \cite{delinker}, a machine learning framework operating on molecular graphs, was proposed to incorporate such structural information. While the results of the method arguably constituted a breakthrough in machine learning-based fragment linking, unfortunately, the method employs a problematic set of coordinates. 

The benefits of a maximally decoupled coordinate system
are well known in computational chemistry and biophysics. For instance, they have turned out especially crucial in the calculation of configurational entropy \cite{killian_extraction_2007, killian_configurational_2009, Hnizdo2010, baron_estimating_2006, MIST, fleck_parent:_2016, Fleck2019, Numata_balanced} from molecular mechanics simulations. 
Here, in Cartesian coordinates, the adequate sampling of the soft constraints present in rigid chemical groups (e. g. a methyl group) quickly becomes computationally prohibitive with increasing molecular size. Recently, similar insights have led to a mathematically rigorous separation of the placeholder atom from the physical partition function in alchemical free energy simulations \cite{fleck_dummy_atoms}. Coordinate systems based on bonds, bond angles and torsions feature decoupling naturally by accommodating the molecular topology made up from rigid bonds. Here, we apply such a bond-angle-torsion (BAT) coordinate system in order to specify the relative orientation of the molecular fragments to be linked. As opposed to the coordinates applied in the investigated previous work \cite{delinker}, the coordinate system proposed in this article is mathematically well-defined, complete and geometrically decoupled. Here, geometrically decoupled denotes the absence of mathematical coordinate constraints \cite{fleck_dummy_atoms}, which constitutes a necessary criterium for the dynamical decoupling mentioned above in the context of configurational entropy.

\section{Methods}

In this article, we propose a beneficial choice of a relative coordinate system for linking pairs of molecular fragments by specifying their relative orientation in 3D space. In order to benchmark our proposed coordinate system, the details of which will be given below, we follow DeLinker
\cite{delinker}, a recent pioneering machine learning framework for linking molecular fragments.

\subsection{The framework}
Here, a brief summary of the DeLinker \cite{delinker} framework is given. For further details, the interested reader is referred to said publication \cite{delinker} as well as references therein (in particular \cite{Liu_2019, Jin_2018, ggnn}).

On the top level, DeLinker \cite{delinker} is embedded in a Variational Autoencoder (VAE) \cite{vae} framework; Linker generation is performed by seeding the latent variables of the decoder of the VAE. First, the graph representation of the fragment pair is encoded by a standard gated graph neural network \cite{ggnn}. Then, a set of atoms is initialized in order to serve as expansion nodes for linking a pair of fragments. The maximum linker length is given by the chosen size of this set of expansion nodes. For each atom, a multidimensional hidden state is drawn from standard normal distributions. The atom type of the respective node is derived via sampling from a learned mapping from the hidden state to a Boltzmann distribution (i.e. softmax or, termed more precisely, softargmax). 

Successively, the fragments are linked by attaching atoms from the set of expansion nodes in an iterative manner. Following the breadth-first paradigm, a first-in-first-out queue is initialized with two exit atoms (Figure \ref{fig:BAT}), one per fragment. The focused node, i. e. the first  node in the queue, forms covalent bonds to candidate atoms, which are themselves added to the queue upon bond formation. The selection process of the focused node is terminated upon forming a bond to a special stop node. Then, the focused node is removed from the queue and the next atom in turn initiates bond formation. Nodes become closed once they are focused, which means they are no longer considered as candidate nodes for bond formation throughout the entire generation process. The bond selection procedure is accomplished by computing feature vectors between the focused atom and all candidate atoms. These feature vectors comprise both atom types, their hidden states as well as their graph distance. Furthermore, the bond formation process takes as an input the average of the hidden states across all nodes during node initialization (i. e. iteration zero) as well as at the current iteration. The current iteration number is supplied in addition. Importantly, the feature vector is augmented with coordinates specifying the relative orientation between the fragments, the details of which will be given below. The actual chosen bond and its type (single, double or triple) are sampled from Boltzmann distributions, which are based on learned mappings from the feature vectors and are masked with valency constraints \cite{Liu_2019}.

After each bond formation, the hidden states of all nodes are updated with respect to the newly formed graph topology. This update is performed via a standard gated graph neural network \cite{ggnn}, taking as input the initial states of all nodes. Starting from the states of the nodes at iteration zero, instead of the current state, prevents the network from learning assembly pathways \cite{Liu_2019}.

The generative process ends when the first-in-first-out queue is empty. Note that the described procedure does not prevent the generation of unlinked fragments, which, however, constitutes the only mechanism leading to invalid molecules as an outcome.

Fragment linking constitutes a multimodal problem. Two fragments can be linked in many different ways. Inspired by Jin et al. \cite{Jin_2018}, a low-dimensional latent vector $z$ is introduced in order for the model to accommodate this one-to-many mapping. To avoid difficulties known from Computer Vision \cite{bicycle_gan}, during training, $z$ is derived from the encoding of the ground-truth linked fragments. In this manner, the model is encouraged to pay attention to the latent vector. Furthermore, $z$ is regularized via the KL-divergence to the standard normal distribution.

\subsection{Relative coordinates}
In this section, we derive our proposed relative coordinate system for linking molecular fragments. The following derivation follows the bond-angle-torsion \cite{killian_extraction_2007, potter_coordinate_2002, chang_calculation_2003, herschbach_molecular_1959, pitzer_energy_1946, go_use_1976, parsons_practical_2005} coordinate formalism, which closely relates to the z-matrix representation \cite{gordon_approximate_1968, parsons_practical_2005}. 

\begin{figure}
	\centering
	\includegraphics[width=.65\linewidth]{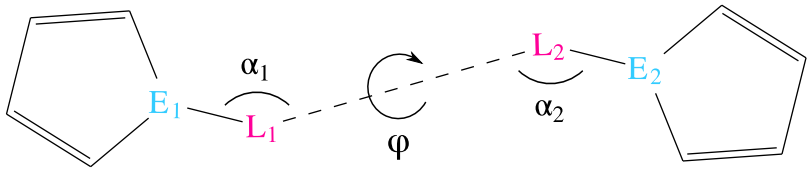}
	\vspace{0.3cm}
	\caption{\textbf{Bond-angle-torsion (BAT) coordinates.} For specifying the relative orientation of two molecular fragments, we first define the exit atom E$_1$ of the first fragment as the atom which is connected to the initial linker atom L$_1$. E$_2$ and L$_2$ are defined analogously for the second fragment. Two of the BAT coordinates are constituted by the exit vectors $|\vec{b_1}|=|\vec{r}_{E_1\rightarrow L_1}|$ and $|\vec{b_3}|=|\vec{r}_{L_2\rightarrow E_2}|$, bridging the exit atoms to the initial linker atoms. A third BAT coordinate is defined by the length of the pseudo-bond $|\vec{b_2}|=|\vec{r}_{L_1\rightarrow L_2}|$. The angles between $\vec{r}_{E_1\rightarrow L_1}$ and $\vec{r}_{L_1\rightarrow L_2}$ as well as between $\vec{r}_{L_1\rightarrow L_2}$ and $\vec{r}_{L_2\rightarrow E_2}$, termed $\alpha_1$ and $\alpha_2$ respectively, yield another two BAT coordinates. In order to obtain six BAT coordinates, a well-known number for specifying relative molecular orientations, we add the dihedral angle $\phi$ of the atom chain E$_1$-L$_1$-L$_2$-E$_2$.}
	\label{fig:BAT}
\end{figure}

Referring to Figure \ref{fig:BAT}, let us introduce the following nomenclature. The atoms of the fragments to which the linker is covalently bound are referred to as exit atoms; E$_1$ and E$_2$ for the two fragments, respectively. The linker atoms L$_1$ and L$_2$ are attached to E$_1$ and E$_2$ via a rotable bond. 
Naturally for the problem of linking molecular fragments, we are not concerned with the global position and orientation of the fragment pair as a whole. Rather, we are interested in the position and orientation of the fragments relative to each other. Thus, without loss of generality, exit atom 1 (E$_1$) is positioned at the
origin of the Cartesian coordinate system, linker atom 1 (L$_1$) on the x-axis and linker atom 2 (L$_2$) in the x-y plane. Then, these anchored Cartesian coordinates \cite{potter_coordinate_2002, chang_calculation_2003} are given as follows:
\begin{align} 
\vec{r}_{E_1}^{\phantom{x}\intercal} &=  (0,0,0) \nonumber\\ 
\vec{r}_{L_1}^{\phantom{x}\intercal} &=  (L_1^x,0,0) \nonumber\\ 
\vec{r}_{L_2}^{\phantom{x}\intercal} &=  (L_2^x, L_2^y, 0) \nonumber\\ 
\vec{r}_{E_2}^{\phantom{x}\intercal} &=  (E_2^x, E_2^y, E_2^z)\nonumber\\ \label{eq:int_cart}
\end{align}
In order to achieve ease of notation, we constitute the following definitions.
\begin{align} 
\vec{b_1}^{\intercal} &\equiv \vec{r}_{E_1\rightarrow L_1}\equiv\vec{r}_{L_1} - \vec{r}_{E_1} = (L_1^x,0,0)\nonumber\\
\vec{b_2}^{\intercal} &\equiv \vec{r}_{L_1\rightarrow L_2}\equiv\vec{r}_{L_2} - \vec{r}_{L_1} = (L_2^x-L_1^x,L_2^y,0)\nonumber\\
\vec{b_3}^{\intercal} &\equiv \vec{r}_{L_2\rightarrow E_2}\equiv\vec{r}_{E_2} - \vec{r}_{L_2} = (E_2^x-L_2^x,E_2^y-L_2^y,E_2^z)\nonumber\\\label{eq:bat1}
\end{align}
Then, our implemented BAT coordinates are given as:
\begin{align} 
|\vec{b_1}| &= |L_1^x|\nonumber\\
|\vec{b_2}| &= \sqrt{(L_2^x-L_1^x)^2 + (L_2^y)^2}\nonumber\\
|\vec{b_3}| &= \sqrt{(E_2^x-L_2^x)^2 + (E_2^y-L_2^y)^2 + (E_2^z)^2}\nonumber\\
\alpha_1 &= \arccos\left(\frac{\vec{b_1}^{\intercal}\cdot\vec{b_2}}{|\vec{b_1}||\vec{b_2}|}\right)\nonumber\\
\alpha_2 &= \arccos\left(\frac{\vec{b_2}^{\intercal}\cdot\vec{b_3}}{|\vec{b_2}||\vec{b_3}|}\right)\nonumber\\
\phi &= \arctan2\left(\frac{|\vec{b_2}|[\vec{b_1}^{\intercal}\cdot(\vec{b_2}\times\vec{b_3})]}
{|\vec{b_1}\times\vec{b_2}||\vec{b_2}\times\vec{b_3}|},
\frac{(\vec{b_1}\times\vec{b_2})^{\overset{\intercal}{\phantom{.}}}\cdot(\vec{b_2}\times\vec{b_3})}
{|\vec{b_1}\times\vec{b_2}||\vec{b_2}\times\vec{b_3}|}\right)\nonumber\\\label{eq:bat2}
\end{align}
Matrix multiplications and outer vector products are denoted by $\cdot$ and $\times$, respectively.
The back-transformation from BAT (Equation \ref{eq:bat2}) to anchored Cartesian coordinates \cite{potter_coordinate_2002, chang_calculation_2003} (Equation \ref{eq:int_cart}), demonstrating the completeness of our proposed BAT coordinate system, is given in the Appendix. In this context, note that we do not explicitly feed the model the lengths of the exit vectors, i. e. $|\vec{b_1}|=|\vec{r}_{E_1\rightarrow L_1}|$ and $|\vec{b_3}|=|\vec{r}_{L_2\rightarrow E_2}|$. As physical bond lengths, their values hardly vary in comparison to the pseudo-bond length $|\vec{b_2}|=|\vec{r}_{L_1\rightarrow L_2}|$, which dominates the geometry. Thus, they barely carry any useful information for the model.

One topic is left to be discussed. In contrast to bond angles, i. e. $\alpha_1$ and $\alpha_2$ in the present case, dihedral angles are periodic. This means $\alpha_1, \alpha_2 \in [0,\pi]$, but $\phi \in [0,2\pi)$. Periodicity poses a challenge for machine learning algorithms, since the implied distance metric does not correctly reflect the underlying physical geometry. 

\begin{figure}  
\centering
     \begin{subfigure}[b]{0.3\textwidth}
         \centering
         \includegraphics[width=\textwidth]{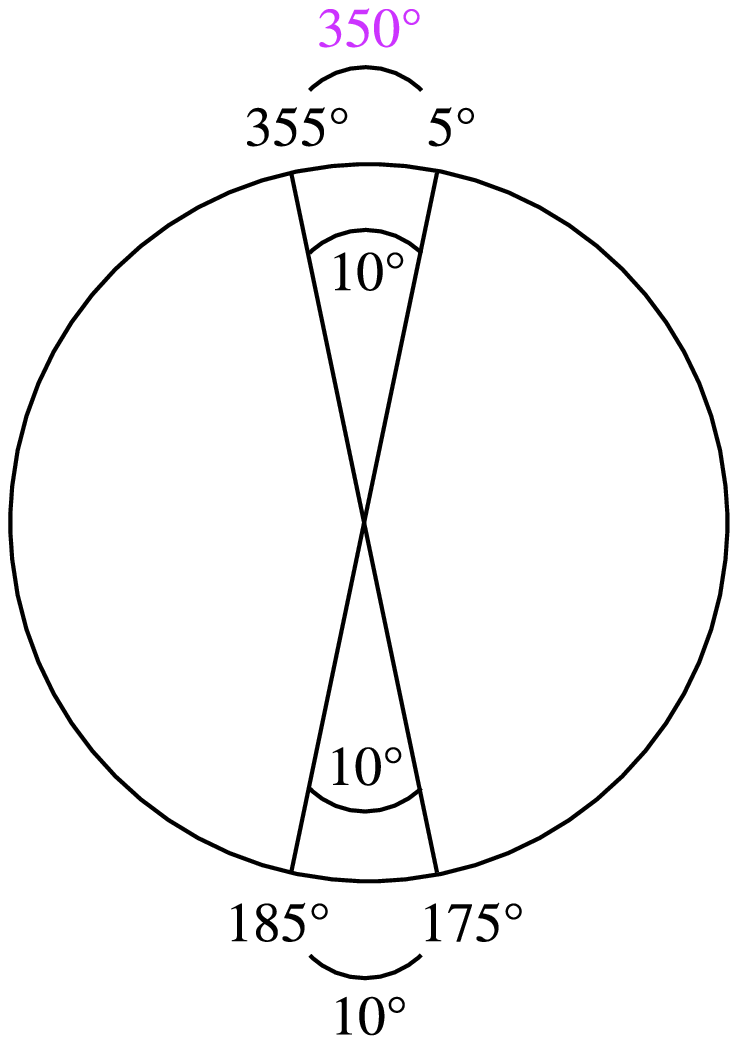}
         \caption{}
     \end{subfigure} 
     \hspace{8ex}
     \begin{subfigure}[b]{0.31\textwidth}
         \centering
         \includegraphics[width=\textwidth, trim = 0cm -1.3cm 0cm 0cm]{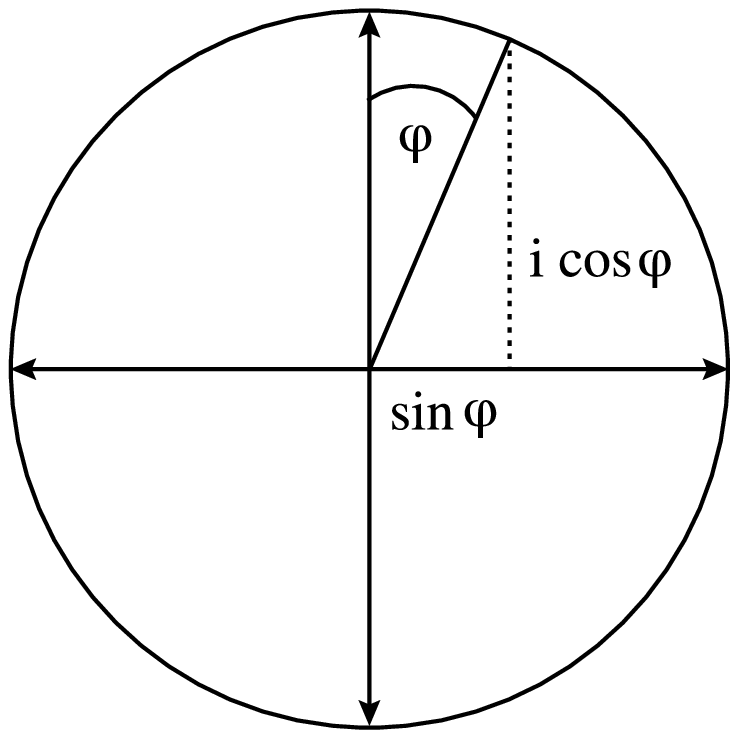}
         \caption{}
     \end{subfigure}
     	\caption{\textbf{Treating periodicity. a)} shows an illustration
     	of the problematic distance metric. Both the top and bottom
     	angles measure 10 degrees geometrically. For the bottom angle,
     	this is correctly reflected in its numerical value, i. e. $|\Delta \alpha_{\text{bottom}}|=|185\degree-175\degree|=10\degree$. 
     	For the top angle, however, the numerical value deviates from 
     	the underlying geometry due to the periodic discontinuity at 
     	$0\degree\,\hat{=}\,360\degree$. Here, we have $|\Delta 
     	\alpha_{\text{top}}|=|5\degree-355\degree|=350\degree$. This 
     	discontinuity of the mapping from geometric angles to their 
     	numerical values poses a challenge for machine learning 
     	algorithms. \textbf{b)} The periodic discontinuity can be 
     	avoided by considering the dihedral an angle in the complex 
     	plane. Supplying both the real and imaginary part, i. e. 
     	$\sin(\phi)$ and $\cos(\phi)$, respectively, dihedral angles are presented to the model in continuous form and without loss of information.}
	\label{fig:circ}
\end{figure}
As done commonly (see e. g. \cite{cosine1, cosine2, cosine3}), we feed the model the sine and cosine of $\phi$, instead of the plain dihedral angle value. One way to motivate this treatment states as follows. There is a periodic discontinuity [see Figure \ref{fig:circ}(a)] at $0\degree\,\hat{=}\,360\degree$ in the mapping of the underlying geometry to its numerical dihedral angle value. A solution is inspired by the complex unit circle [Figure \ref{fig:circ}(b)]: One maps the dihedral $\phi$ to $\exp[i (\pi/2-\phi)]= \sin(\phi)+i\cos(\phi)$. The machine learning model is fed the real and imaginary part of this transformation, i. e. both $\sin(\phi)$ and $\cos(\phi)$. These angular functions, as a representation of the periodic dihedral angle $\phi$, are both continuous as well as faithful with respect to the underlying geometry. Additionally, they naturally map to the interval $[-1,1]$, which constitutes another desirable property. Note that the transformation $\exp[i (\pi/2-\phi)]$ was chosen over just $\exp (i\phi)$ in order to match the complex unit circle with the canonical definition of dihedral angles [Figure \ref{fig:circ}(b)]. However, as the order of the input features is irrelevant for the machine-learning algorithm, this choice of swapping sine with cosine is arbitrary. 

\subsection{Data preparation}
\label{sec:data_prep}
Following DeLinker \cite{delinker}, we are working with two datasets, a selected \cite{ZINC250k} set of 250 000 ZINC \cite{ZINC} compounds as well as CASF-2016 (i. e., the PDBbind core set) \cite{CASF-2016}. A major difference between these two sets is the origin of the 3D structural information. While for CASF, the structures stem from 285 high-quality crystal structures of ligands bound to proteins, the ZINC structures are generated \textit{in silico} using RDKit \cite{RDKit}, as further detailed below. 

For both data sets, preparation is performed in the same manner and sketched as follows. The ligands are split into three parts, i. e. two fragments and the corresponding linker. Cuts were performed on acyclic single bonds outside of functional groups \cite{acyclic_cuts}. Triplet splits for which either of the three components is unrealistically small, or the linker is inflated with respect to the fragments, were removed. The remaining triplet splits were filtered further by molecular graph (2D) properties, in particular synthetic accessibility (SA) \cite{SA}, ring aromaticity as well as pan-assay interference compounds (PAINS) \cite{PAINS}. This procedure yielded $\sim$420 000 triplet splits for the ZINC data set and 309 triplet splits for CASF.

Training is performed purely on the ZINC data set. 400 compounds each were randomly selected from the whole ZINC set and held back as a validation and test set. The CASF data is used solely as a test set for evaluation.

As outlined above, unlike for CASF, where experimental 3D structures are available, the ZINC 3D coordinates need to be generated. This is accomplished using RDKit \cite{RDKit} via employing the Merck molecular force field (MMFF) \cite{MMFF94, MMFF}. The DeLinker \cite{delinker} source code ships without the 3D structures of the training set (obviously due to storage space reasons). For calculating our proposed relative coordinates of the training set, we reproduced the 3D structures. Referring to Figure \ref{fig:BAT}, DeLinker \cite{delinker} uses the distance from E$_1$ to E$_2$ as well as the angle between the exit vectors (i. e. $\vec{r}_{E_1\rightarrow L_1}$ and $\vec{r}_{E_2\rightarrow L_2}$) as relative coordinates. In order to ensure exact reproduction, we recalculated these relative coordinates alongside the proposed BAT coordinates and compared them to the data shipped with DeLinker \cite{delinker}. Our comparison demonstrated an exact match. For further details, the interested reader is referred to the DeLinker publication \cite{delinker}.

\subsection{Training}
The model is trained under a VAE framework over 10 epochs, exclusively on the ZINC training set, using the Adam optimizer with a learning rate of $10^{-3}$ and a minibatch size of 16. The encoding of the nodes hidden states as well as the latent vector $z$ encoding the ground-truth molecule are regularized via KL loss to follow standard normal distributions. A two-fold cross-entropy reconstruction loss is applied, one part measuring the error in the prediction
of the atom types, the second part judging the sequence of bond-formation steps in order to reconstruct the ground truth molecule. Further details can be found in the DeLinker publication \cite{delinker} as well as in Liu et al. \cite{Liu_2019}. 

\subsection{Evaluation}
\label{sec:eval}
For each of the fragment pairs in the triplet cuts of the test sets, i. e. 400 in the case of ZINC and 309 for CASF, 250 linkers are generated. This amounts to 100 000 and 77 250 linked molecules generated, respectively. We apply the standard metrics for molecular generation tasks, i. e. fractional validity, uniqueness and novelty. Validity is assessed by RDKit \cite{RDKit} being able to parse the generated SMILES \cite{SMILES1} strings as connected molecules. Uniqueness is calculated by the cardinality of the (unique) set of generated molecules divided by the total number generated. Novelty describes the fraction of generated molecules which are not contained in the training set.

Furthermore, we assess the generated molecules via the 2D filtering metrics of subsection \ref{sec:data_prep}. SA \cite{SA} is a measure for the difficulty of physically synthesizing the molecule in the laboratory. Ring aromaticity amounts to the properties of a molecule constituting a drug. PAINS \cite{PAINS} assesses the reactivity of a compound by performing a knowledge-based analysis on its substructures. 

Arguably the most significant estimator of the impact of our coordinate system is the models capability to recover the ground-truth linker from the original triplet cut. This statement is motivated by the fact that the structural information provided to the generation process is derived from the respective ground-truth linker. In this manner, the model is conditioned to reproduce the ground-truth linker more frequently.

\subsection{Information-theoretical analysis}
Input features, referring to the coordinates in the present case, should be uncorrelated, i. e. decoupled, in order to assure efficient learning for the model. Therefore, the mutual information between the input features should be low \cite{brown_mutual}. In order to investigate said decoupling, we calculate pairwise mutual information values, given as \cite{killian_extraction_2007, killian_configurational_2009, Hnizdo2010, fleck_parent:_2016, Fleck2019, Numata_balanced}
\\
\begin{equation}
I(X,Y) = S(X) + S(Y) - S(X,Y)
\label{MI_1}
\end{equation}
with
\begin{align}
S(X) &= -\sum_i p_i^x \ln\left(\frac{p_i^x}{J_i^x\Delta x}\right)\nonumber\\
S(Y) &= -\sum_j p_j^y \ln\left(\frac{p_j^y}{J_j^y\Delta y}\right)\nonumber\\
S(X,Y) &= -\sum_{i,j} p_{i,j}^{x,y} \ln\left(\frac{p_{i,j}^{x,y}}{J_{i}^{x}J_{j}^{y}\Delta x\Delta y}\right)\nonumber\\
\label{entropy}
\end{align}
The formulas given in Equation \ref{entropy} refer to discretized differential entropy values. This means the underlying continuous probability densities are approximated by sampling to discrete histogram bins. The indices x and y designate an arbitrary coordinate, i. e. a bond, an angle or a dihedral in the present case. $p_i^x$ denotes the probability for the coordinate $x$ to assume a value in bin $i$. If $N$ samples for coordinate $x$ are taken, and $N_i$ of them fall into bin $i$, then $p_i^x=N_i/N$ (and analogous for $p_j^y$). Furthermore, $p_{i,j}^{x,y}=N_{i,j}/N$ if $N_{i,j}$ of the $N$ samples taken fall into bin $i$ for the $x$-coordinate and bin $j$ for the $y$-coordinate in the according 2D histogram. $\Delta x$ and $\Delta y$ are the widths of the equally spaced bins. $J_i^x$ is the Jacobian of coordinate $x$ in the middle of bin $i$. Using the definition of the marginals
\begin{align}
p_i^x &= \sum_j p_{i,j}^{x,y}\nonumber\\
p_j^y &= \sum_i p_{i,j}^{x,y},\nonumber\\
\end{align}
we can write Equation \ref{MI_1} as 
\begin{align}
\label{MI_2}
I(X,Y) &= \sum_{i,j} p_{i,j}^{x,y} \ln\left(\frac{p_{i,j}^{x,y}/(J_{i}^{x}J_{j}^{y}\Delta x\Delta y)}
{p_{i}^{x}/(J_{i}^{x}\Delta x)*p_{j}^{y}/(J_{j}^{y}\Delta y)}\right)\nonumber\\
&=\sum_{i,j} p_{i,j}^{x,y} \ln\left(\frac{p_{i,j}^{x,y}}{p_{i}^{x}p_{j}^{y}}\right)\nonumber\\
\end{align}
The last equality exhibits interesting features to be discussed. First, the Jacobian determinants have canceled out, meaning the type of degree of freedom has become irrelevant. Furthermore, the bin sizes have vanished. This means that the mutual information calculated from discretized differential entropy
values resembles discrete information in Shannon's \cite{shannon_mathematical_1948} sense.
Note, however, that the mutual information still is dependent on bin sizes; the probabilities $p_{i,j}^{x,y}$, $p_{i}^{x}$ and $p_{j}^{y}$ are affected by this choice. For example, let us denote the limit where the binning becomes binary, meaning the bin sizes are chosen so small that there is either no or only one data point in each bin. Assuming that both the x- as well as the y-marginal take on such a binary form, we can write for $N$ data points:
\begin{align}
\label{MI_3}
I(X,Y) &=\sum_{i,j} p_{i,j}^{x,y} \ln\left(\frac{p_{i,j}^{x,y}}{p_{i}^{x}p_{j}^{y}}\right)\nonumber\\
&=\sum_{p_{i,j}^{x,y} > 0}\frac{1}{N}\ln\left(\frac{1/N}{(1/N)*(1/N)}\right)\nonumber\\
&=\frac{N}{N}\ln\left(N\right)\nonumber\\
&=\ln(N)\nonumber\\
\end{align}
Furthermore, note that we write the equations without the Boltzmann or Gas constant. Spatial entropies, as in the equations \ref{entropy}, do not bear physically meaningful units. The reason is the fact that the semi-classical entropy integral cannot be split in a manner that either the spatial or the momentum entropy bear physically meaningful units \cite{Hnizdo2010}. Upon taking entropy differences, however, the problematic terms cancel out \cite{Hnizdo2010}. This means that the mutual information in equations \ref{MI_1}, \ref{MI_2} and \ref{MI_3} can indeed be multiplied with the Boltzmann or Gas constant to yield physical entropy values. Stated without such a physical constant, the units obtained here are natural units of information (nats, similar to bits, however, using the natural logarithm).

\section{Results and discussion}
As mentioned in section \ref{sec:eval}, arguably the most indicative metric for the quality of the provided coordinates is the rate of recovery of the ground-truth linker, as it quantifies the models capability to follow the supplied information. This metric improves across all test sets, as shown in Table \ref{tab:graph_metrics}, with the most drastic enhancement for ZINC. The coordinate system carries information about the ground-truth linker. With higher quality information supplied, the model is expected to recover the original linker more frequently. This demonstrates that the model takes advantage of the augmented information by following the provided target geometry. Since the generation process is more directed, the uniqueness metrics drops, which serves as another indicator for the quality of the information content in our proposed coordinate system.
\begin{table}
\centering
    \begin{tabular}{ccc|c|c|c||c}

 &&& \multicolumn{1}{c|}{no info} & \multicolumn{1}{c|}{distance only} & \multicolumn{1}{c||}{DeLinker} & \multicolumn{1}{c}{\phantom{ab}BAT\phantom{ab}}\\
 \hline
 \hline
 \rowcolor{Gray}&&ZINC &74.5 &78.3 &79.0 &\textbf{88.3}\\
 \rowcolor{Gray}&&CASF &N/A&N/A&53.7 &\textbf{56.3}\\
 \rowcolor{Gray}&&ZINC 5+ atoms &N/A&N/A&67.0 &\textbf{80.8}\\
 \parbox[t]{10mm}{\centering \multirow{-4}{*}{\cellcolor{Gray}\rotatebox[origin=c]{90}{recovered}}} &\cellcolor{Gray}&\cellcolor{Gray}CASF 5+ atoms &\cellcolor{Gray}N/A&\cellcolor{Gray}N/A &\cellcolor{Gray}29.8&\cellcolor{Gray}\textbf{34.0}\\
\hline 
\parbox[t]{10mm}{\centering \multirow{4}{*}{\rotatebox[origin=c]{90}{novel}}} 
 &&ZINC &36.2 &37.6 &39.5 &\textbf{39.6}\\
 &&CASF  &N/A&N/A&51.0 &\textbf{53.3}\\
 &&ZINC 5+ atoms  &N/A&N/A&\textbf{49.4} &47.9\\
 &&CASF 5+ atoms &N/A&N/A &68.7& \textbf{69.1}\\
\hline 
\parbox[t]{10mm}{\centering \multirow{4}{*}{\rotatebox[origin=c]{90}{valid}}} 
 &&ZINC &97.0&\textbf{98.6} &98.4 &98.2\\
 &&CASF  &N/A&N/A&\textbf{95.5} &94.5\\
 &&ZINC 5+ atoms  &N/A&N/A&98.1 &\textbf{98.3}\\
 &&CASF 5+ atoms &N/A&N/A &\textbf{94.7}& 93.1\\
\hline
 \parbox[t]{10mm}{\centering \multirow{4}{*}{\rotatebox[origin=c]{90}{unique}}} 
 &&ZINC &\textbf{51.2} &47.3 &44.2 &37.6\\
 &&CASF  &N/A&N/A&\textbf{51.9} &47.1\\
 &&ZINC 5+ atoms  &N/A&N/A&\textbf{61.0} &52.7\\
 &&CASF 5+ atoms &N/A&N/A &\textbf{72.9}& 66.3\\
\hline
\hline
 \parbox[t]{10mm}{\centering \multirow{4}{*}{\rotatebox[origin=c]{90}{\parbox{1cm}{\centering pass all 2D filters}}}} 
 &&ZINC &89.9 &90.2 &89.8 &\textbf{90.5}\\
 &&CASF  &N/A&N/A&\textbf{81.4} &80.4\\
 &&ZINC 5+ atoms  &N/A&N/A&84.1 &\textbf{85.5}\\
 &&CASF 5+ atoms &N/A&N/A &\textbf{71.7}& 70.3\\
\hline
\parbox[t]{10mm}{\centering \multirow{4}{*}{\rotatebox[origin=c]{90}{\parbox{1cm}{\centering pass ring filter}}}} 
 &&ZINC &95.2 &94.5 &94.8 &\textbf{95.5}\\
 &&CASF &N/A&N/A&N/A &92.5\\
 &&ZINC 5+ atoms  &N/A&N/A&N/A &92.2\\
 &&CASF 5+ atoms &N/A&N/A &N/A& 87.2\\
\hline
\parbox[t]{10mm}{\centering \multirow{4}{*}{\rotatebox[origin=c]{90}{\parbox{1cm}{\centering pass SA filter}}}} 
 &&ZINC &95.1 &\textbf{95.5} &95.3 &\textbf{95.5}\\
 &&CASF  &N/A&N/A&N/A &85.1\\
 &&ZINC 5+ atoms  &N/A&N/A&N/A &93.4\\
 &&CASF 5+ atoms &N/A&N/A &N/A& 77.6\\
\hline
\parbox[t]{10mm}{\centering \multirow{4}{*}{\rotatebox[origin=c]{90}{\parbox{1cm}{\centering pass PAINS filter}}}} 
 &&ZINC &97.8 &\textbf{98.4} &97.9 &98.2\\
 &&CASF &N/A&N/A&N/A &98.0\\
 &&ZINC 5+ atoms &N/A&N/A&N/A &97.7\\
 &&CASF 5+ atoms &N/A&N/A &N/A& 97.7\\
\hline
\hline
 \end{tabular}
 \vspace{3mm}
\caption{\textbf{Graph metrics.} 2D quality criteria are compared for molecules generated using BAT coordinates versus the values from the DeLinker \cite{delinker} publication. Values which are not given in the DeLinker \cite{delinker} publication are marked via an "N/A" entry. For the ZINC dataset, the DeLinker \cite{delinker} ablation study values are included. The two 5+ test sets at the bottom constitute subsets of target linkers with at least 5 atoms in length. Best in category (row) values have been printed in bold font.}
\label{tab:graph_metrics}
\end{table}

\begin{figure}
	\centering
	\begin{tabular}{|cc|}
	\hline
	&\\
	\phantom{ab}\includegraphics[width=.45\linewidth]{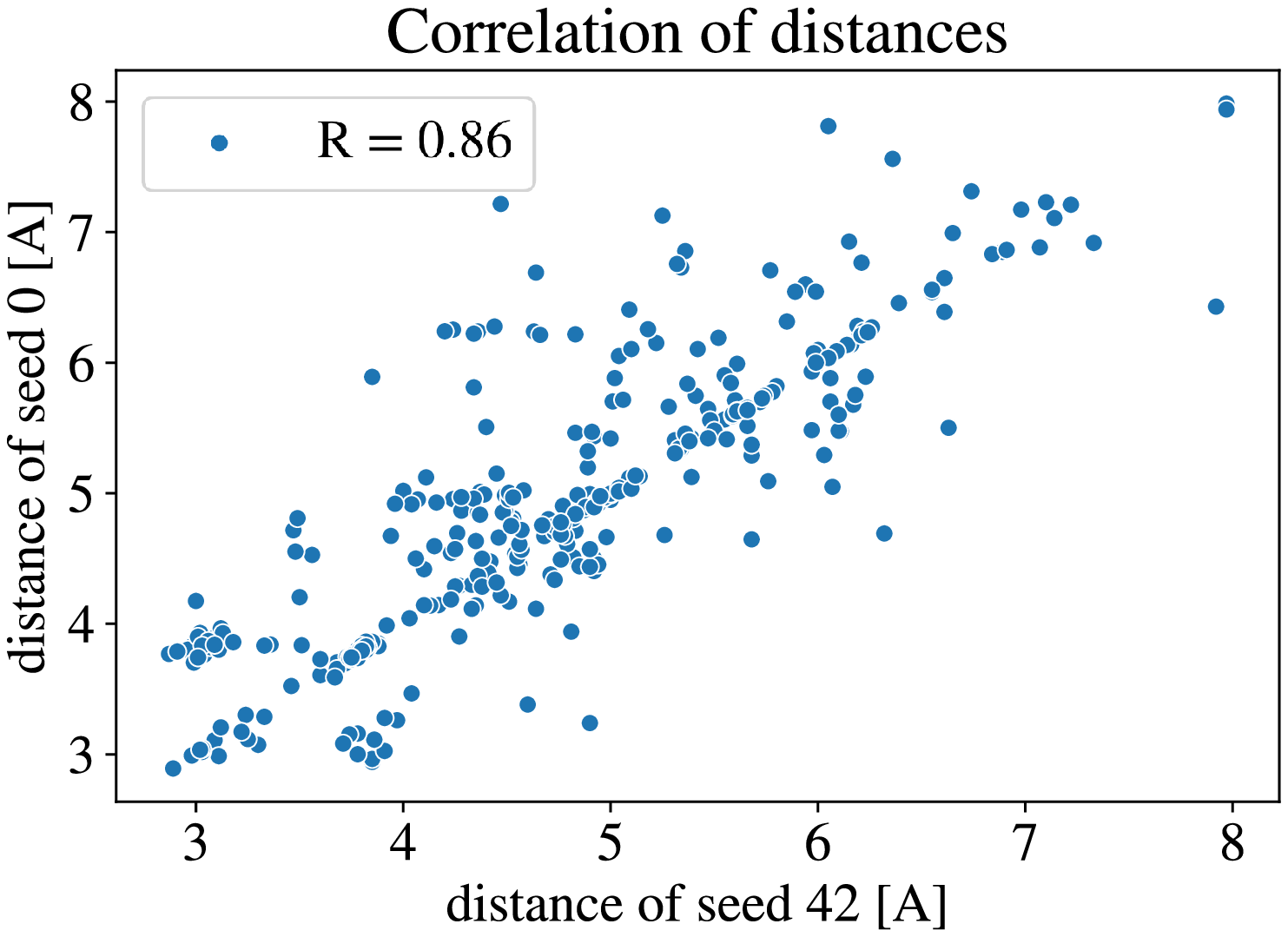}
	&\includegraphics[width=.45\linewidth]{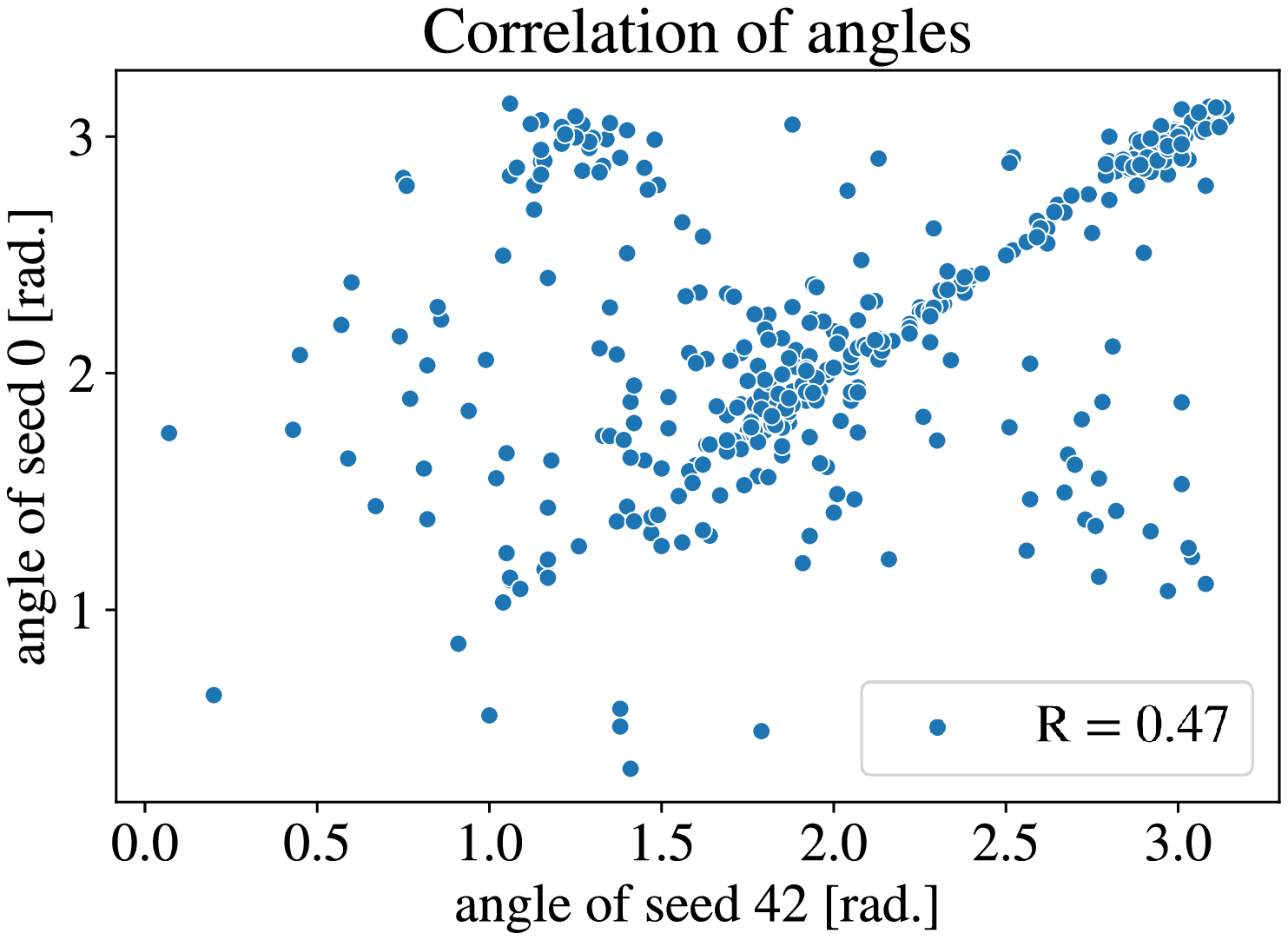}\phantom{ab}\\
	&\\
	\hline
	\end{tabular}
	\vspace{1mm}
	\caption{\textbf{Robustness of DeLinker coordinates.} The plots compare two random seeds used for generating conformers for the test set. The distance as well as the angle of the DeLinker coordinate set are plotted. The distances demonstrate a somewhat adequate robustness between random seeds, with Pearson correlation of 0.86 between seeds. For the angles, we observe a considerable fraction of outliers, decreasing the correlation to 0.47 only. The robustness to altering random seeds is proposed as a proxy for the quality of the provided structural information.}
	\label{fig:their_coordinate_quality}
\end{figure}

Considering the ablation study of DeLinker \cite{delinker} on the ZINC test set, the bulk of the geometric information obviously is contained in the distance coordinate. Compared to providing no relative coordinates at all, the distance takes the recovery rate from 74.5 to 78.3 percent. Adding the angle coordinate, i. e., considering the complete DeLinker \cite{delinker} coordinate set, yields an additional gain of comparatively low 0.7 percent. Figure \ref{fig:their_coordinate_quality} provides insight on this outcome.
When using a different random seed for RDKit to generate conformers, the DeLinker \cite{delinker} distances are preserved rather robustly with a Pearson correlation of 0.86. The angles, on the other hand, show significant deviations. Their Pearson correlation to the angles of the conformers generated with the previous random seed is given as a relatively low value of 0.47. This indicates that the reliability of the angular information is low compared to the distance information, which attributes to only minor gains when feeding them to the model.

In order to further investigate the information content in the relative coordinates proposed here, analogue ablation studies were performed for the BAT coordinates. Table \ref{tab:our_ablation} lists the results. 
\begin{table}
\centering
    \begin{tabular}{ccc||c|c|c||c}

 &&& \multicolumn{1}{c|}{no info} & \multicolumn{1}{c|}{distance only} & \multicolumn{1}{c||}{distance and angles} & \multicolumn{1}{c}{\phantom{ab}BAT\phantom{ab}}\\
 \hline
 \hline
 \parbox[t]{2mm}{\multirow{8}{*}{\rotatebox[origin=c]{90}{ZINC}}} 
 &&\cellcolor{Gray}recovered &\cellcolor{Gray}74.5 &\cellcolor{Gray} 84.5&\cellcolor{Gray}87.0 &\cellcolor{Gray}\textbf{88.3}\\ &&novel &36.2 &\textbf{40.6} & 38.6&39.6\\
 &&valid &97.0& 97.5 &\textbf{98.2} & \textbf{98.2}\\
 &&unique &\textbf{51.2} & 44.8&37.8 &37.6\\
 \cline{3-7}
 &&pass all 2D filters &89.9 & 89.8& \textbf{90.5}&\textbf{90.5}\\
 &&pass ring filter &95.2 & 94.5& \textbf{95.6}&95.5\\
 &&pass SA filter &95.1 & 95.1& 94.9&\textbf{95.5}\\
 &&pass PAINS filter &97.8 &98.0 & \textbf{98.2} &\textbf{98.2}\\
\hline
\hline
 \end{tabular}
 \vspace{4mm}
\caption{\textbf{BAT ablation study.} The same metrics as in Table \ref{tab:graph_metrics} are investigated and the impact of the different types of coordinates is dissected. As was the case for the DeLinker \cite{delinker}  coordinates, the 2D filter and the valid criteria are generally high. Referring to the recovered metric, similar to Table \ref{tab:graph_metrics}, the bulk of our information is carried in the distance coordinate, albeit the improvement appears significantly more distinct. Furthermore, the unique metric continues to behave opposite to the recovery metric, supporting the hypothesis of the coordinate information providing valuable guidance for the model.}
\label{tab:our_ablation}
\end{table}

\begin{figure}
	\centering
	\begin{tabular}{|cc|}
	\hline
	&\\
	\phantom{ab}\includegraphics[width=.45\linewidth]{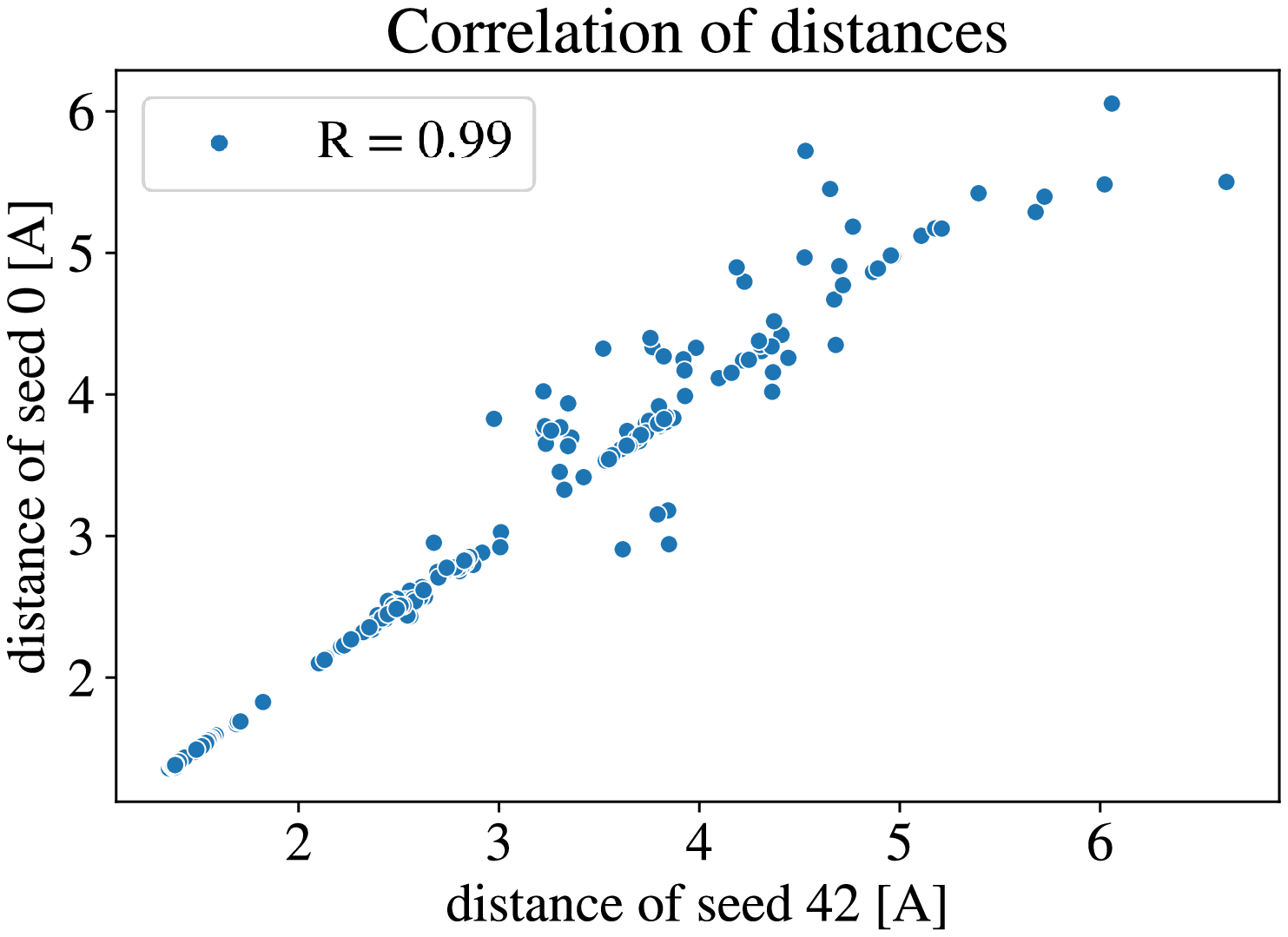}
	&\includegraphics[width=.45\linewidth]{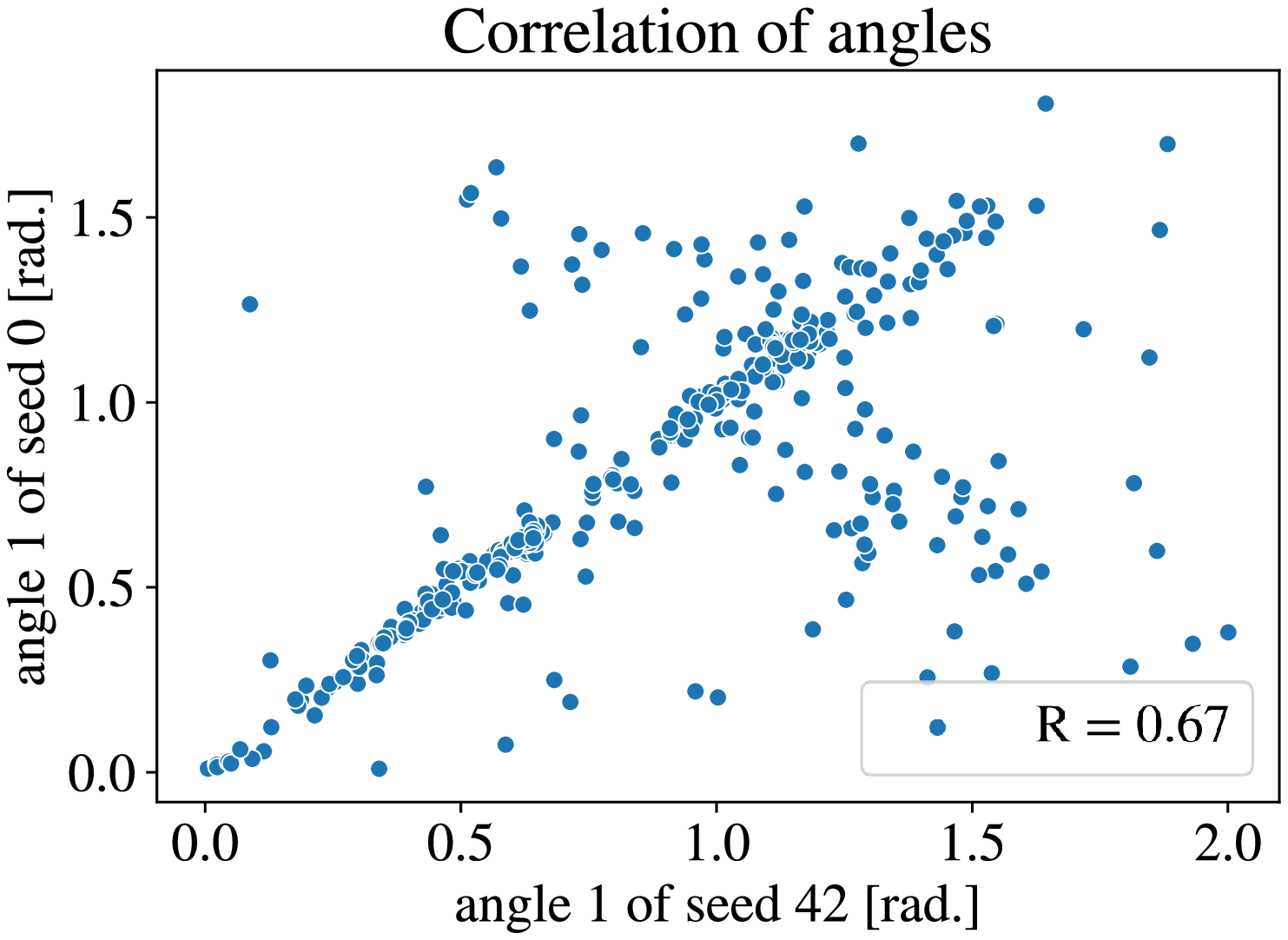}\phantom{ab}\\
	&\\
	\includegraphics[width=.45\linewidth]{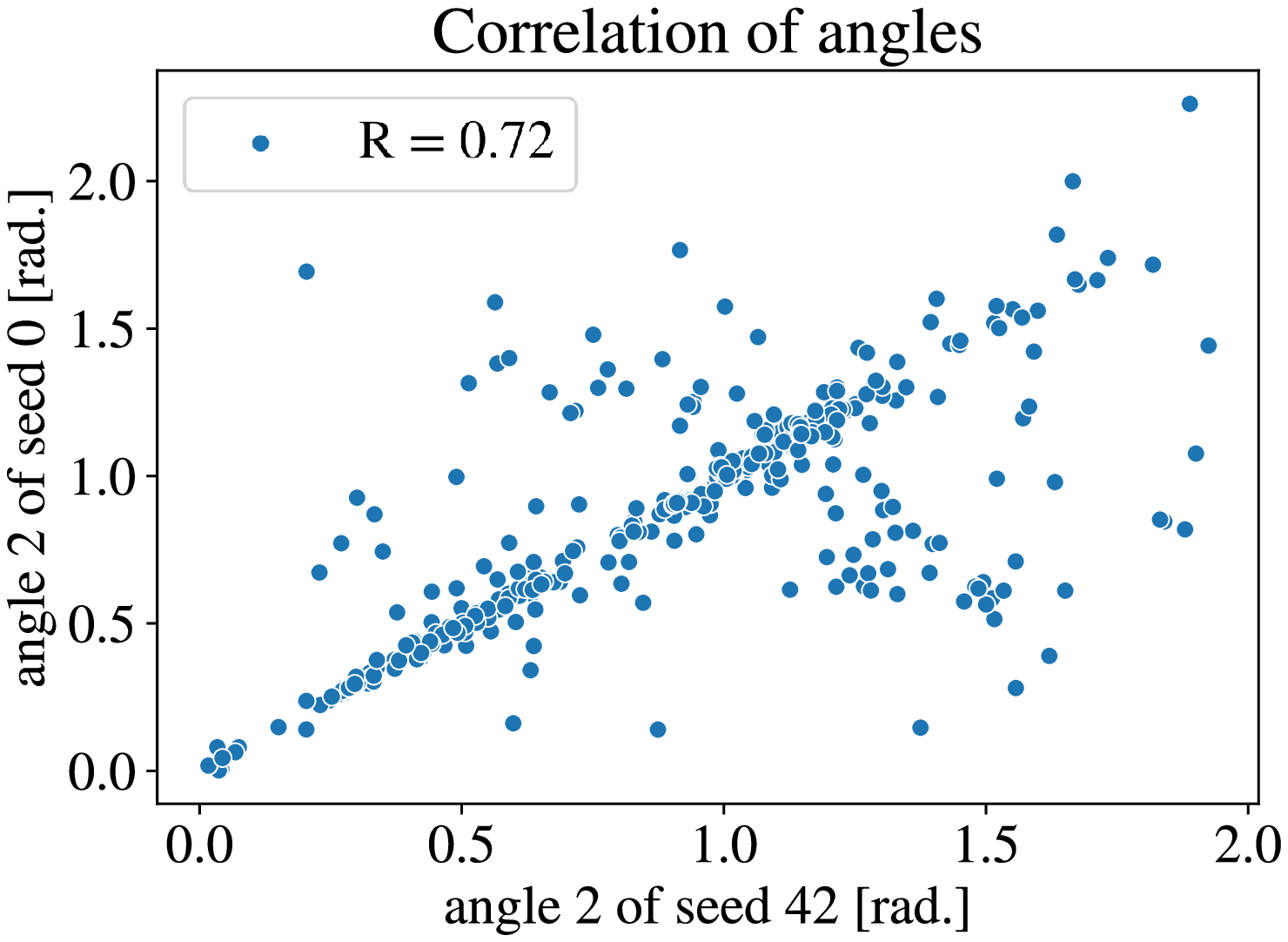}
	&\includegraphics[width=.45\linewidth]{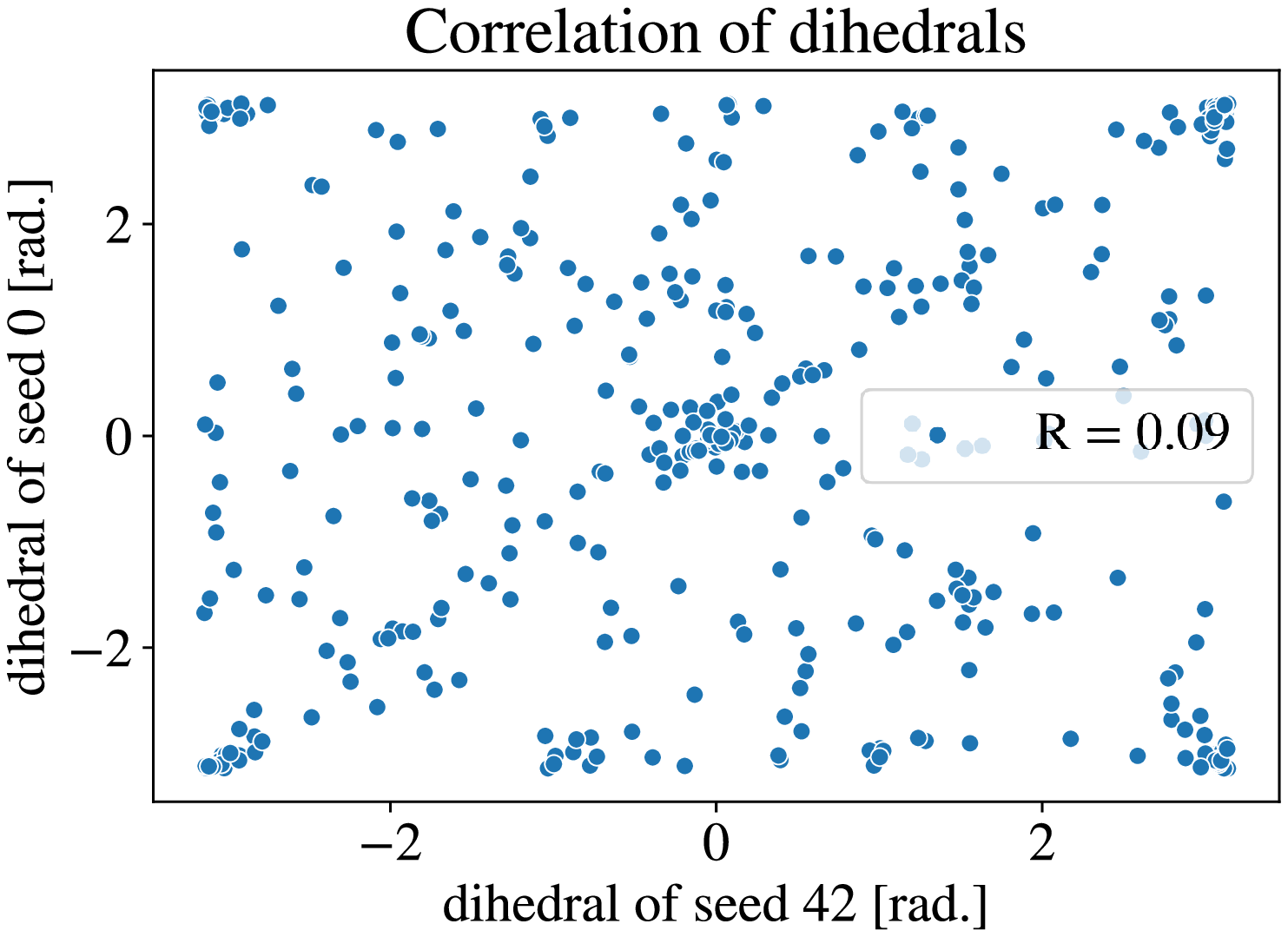}\\
	&\\
	\hline
	\end{tabular}
	\vspace{1mm}
	\caption{\textbf{BAT coordinate quality.} Analogous plot to Figure \ref{fig:their_coordinate_quality} for the proposed BAT coordinates. The robustness to varying random seeds for the distances appears excellent with a Pearson $R=0.99$. Compared to Figure \ref{fig:their_coordinate_quality}, both our angles demonstrate enhanced reliability. Referring to the dihedrals, the estimated information content is considerably low.}
	\label{fig:our_coordinate_quality}
\end{figure}

Similar to the DeLinker \cite{delinker} coordinate system, the bulk of the BAT information is carried in the distance coordinate. However, the BAT distance ($|\vec{r}_{L_1\rightarrow L_2}|$ in Figure \ref{fig:BAT}) takes the recovered metric from 74.5 to 84.5 (Table \ref{tab:our_ablation}), as opposed to only 78.3 (Table \ref{tab:graph_metrics}) for the DeLinker \cite{delinker} distance $|\vec{r}_{E_1\rightarrow E_2}|$. Given the similarity of the two distances, this discrepancy appears rather drastic at a first glance. The reason is argued to lie in the fact that the DeLinker \cite{delinker} distance fails to decouple from the angular and dihedral coordinate systems. For the 6 BAT coordinates in Figure \ref{fig:BAT}, one can vary each of them while keeping the others constant. However, any such variation will change the DeLinker \cite{delinker} distance $|\vec{r}_{E_1\rightarrow E_2}|$. Furthermore, varying the BAT angles or the BAT dihedral will change the DeLinker \cite{delinker} angle. When comparing the distances in Figure \ref{fig:their_coordinate_quality} to those in Figure \ref{fig:our_coordinate_quality}, the effect of this coupling becomes evident. 
The DeLinker \cite{delinker} distance shows a Pearson correlation of 0.86 between coordinates of conformers generated with different random seeds. The decoupled BAT distance, on the other hand, demonstrates excellent robustness to the choice of random seed with a Pearson correlation of 0.99. This explains the arguably drastic improvement of the recovered metric by 10 percent using the BAT distance only.

Feeding the model the BAT angles additionally, we gain another 2.5 percent, as opposed to 0.7 for the DeLinker \cite{delinker} angle. Comparing the correlation plots, BAT provides two angles with a Pearson correlation of 0.67 and 0.72 versus one angle with 0.47 for DeLinker \cite{delinker}; the BAT angles are decoupled from the distance and dihedral system.

Inspecting Figure \ref{fig:our_coordinate_quality}, the dihedral angle information is of low quality. Interestingly, even given this low quality information, the model improves the recovered metric by 1.3 percent. Note that this value exceeds the angular improvement of the DeLinker \cite{delinker} coordinates (0.7) almost by a factor of two. The reason, as above, arguably lies in the decoupling from the distance and angular system; while the information is certainly non-reliable on its own, it is indeed orthogonal, i. e. non-redundant, with respect to the other coordinates. Altogether, the BAT coordinate system takes the recovery metric to 88.3 percent, which is an improvement of 9.3 percent over the DeLinker \cite{delinker} coordinate system.

\begin{figure}
	\centering
	\begin{tabular}{|c|}
	\hline
	\\
	\phantom{ab}\includegraphics[width=.45\linewidth]{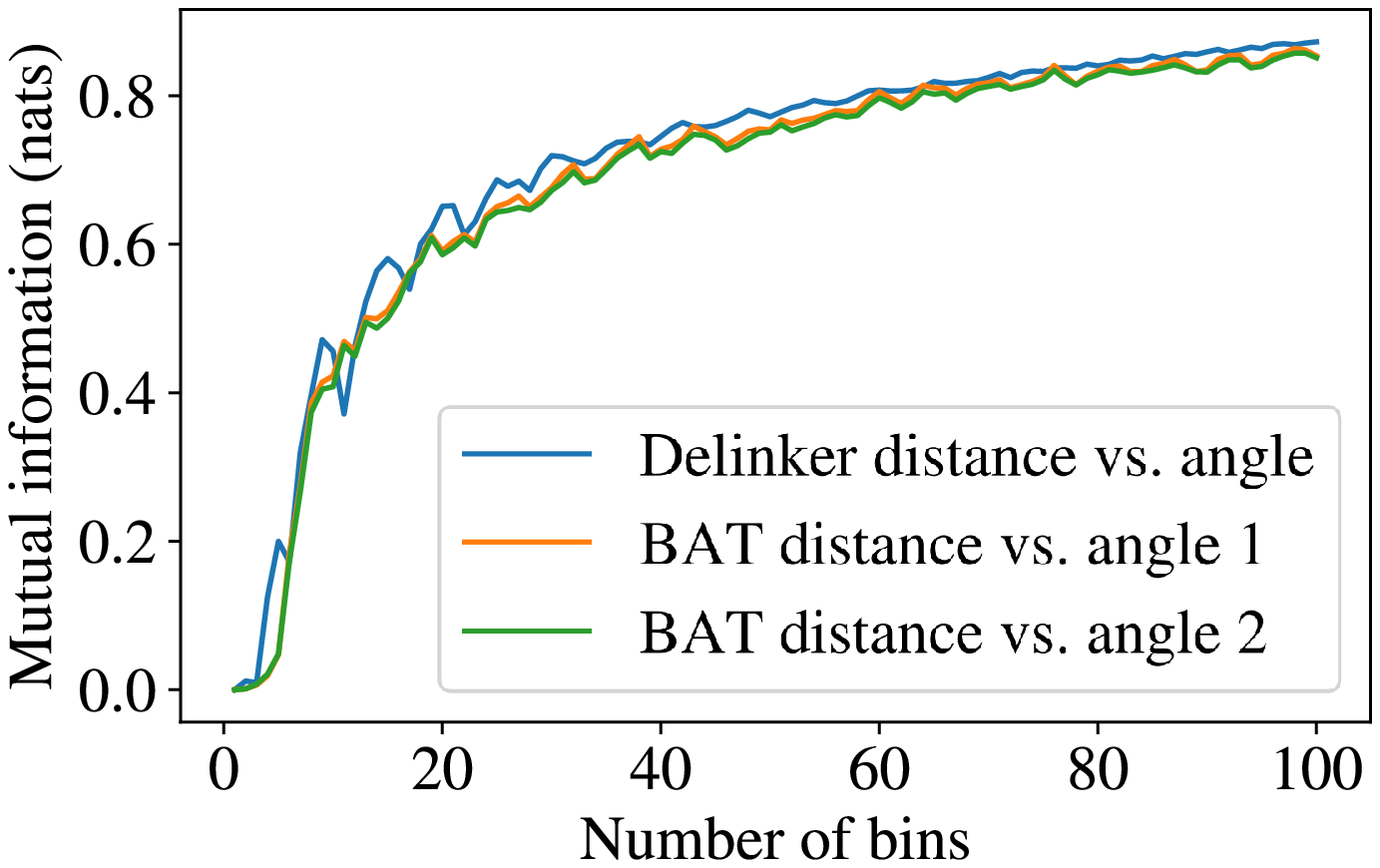}\phantom{ab}\\
	\\
	\includegraphics[width=.45\linewidth]{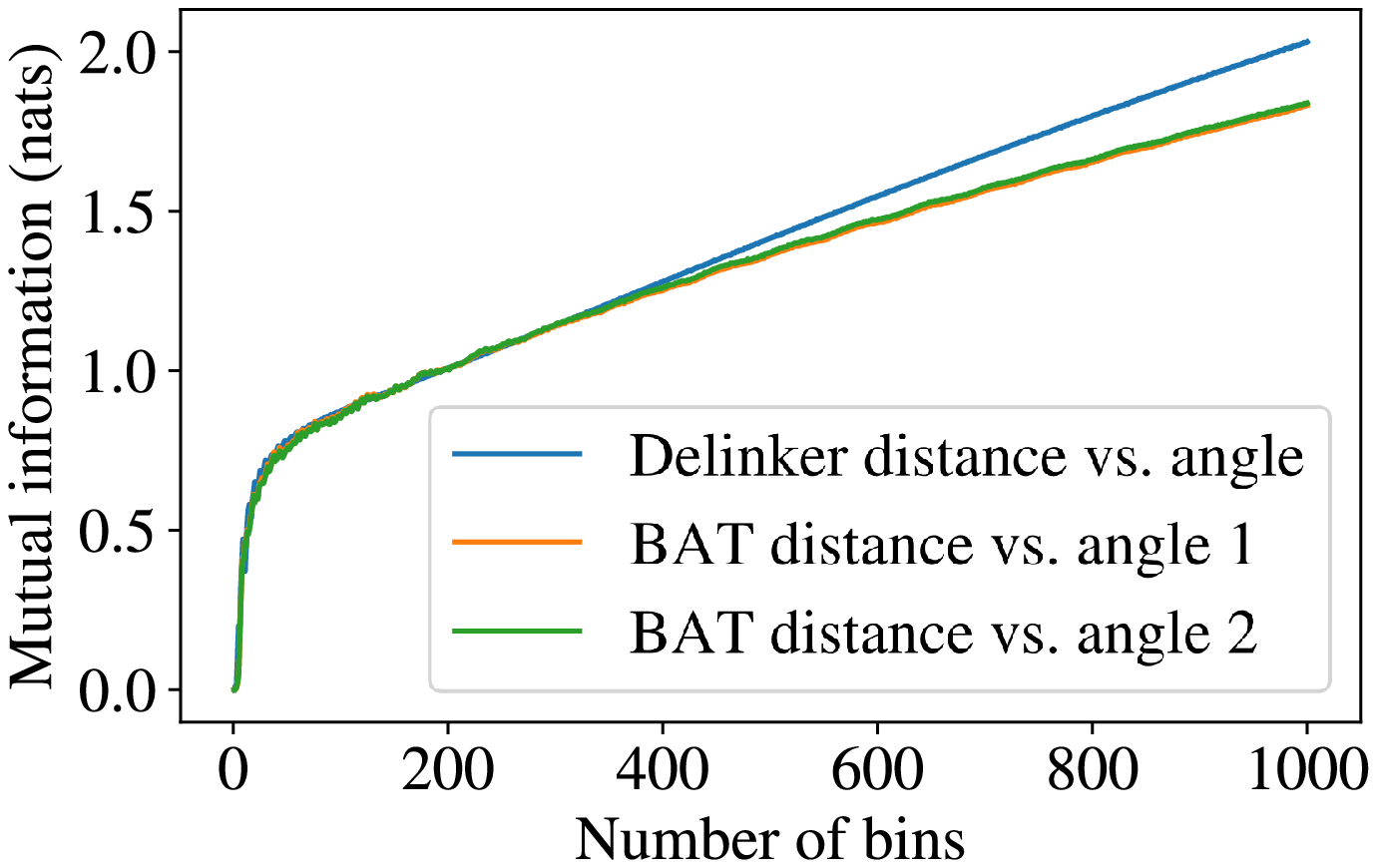}\\
	\\
	\includegraphics[width=.45\linewidth]{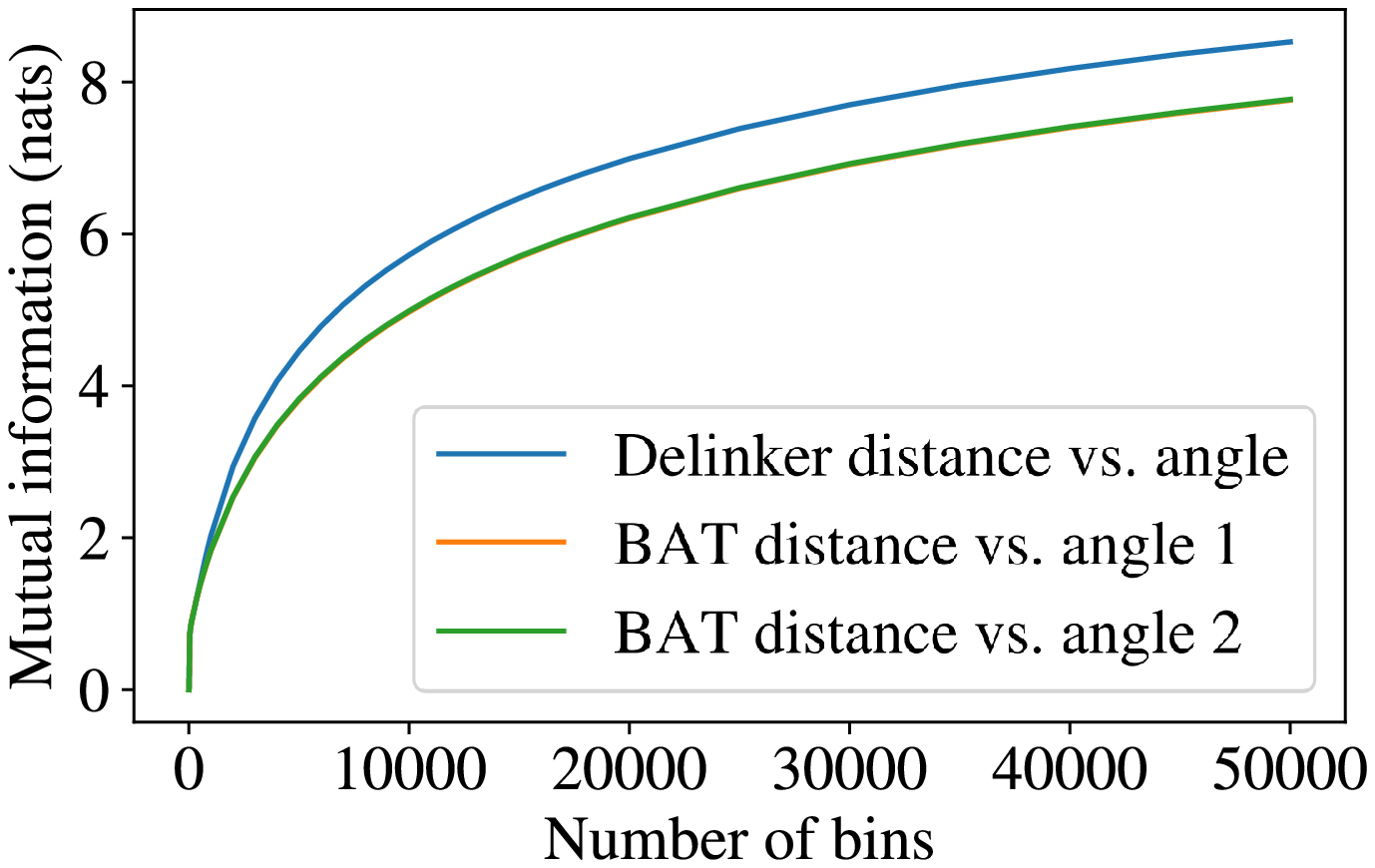}\\
	\\
	\hline
	\end{tabular}
	\caption{\textbf{Mutual information comparison.} The graphs compare the mutual information between the DeLinker \cite{delinker} distance and angle as well as between the BAT distance and bond angles. The training set of ground-truth molecules is analyzed. Since, as discussed in the Methods section, the mutual information is dependent on the bin size, the plots show the mutual information as a function of the granularity in terms of the number of bins used. Different regimes of the number of bins are shown. Note that the mutual information curves for both our angles lie on top of each other. With the exception of rare events in the regime of low number of bins (top graph), our coordinates exhibit lower mutual information, i. e. enhanced decoupling.}
	\label{fig:MI}
\end{figure}

Last but not least, we have performed an information-theoretical analysis comparing the two input coordinate sets. Given the fact that the DeLinker \cite{delinker} coordinates are given as a distance and angle pair, we compare the mutual information using the training set between the Delinker \cite{delinker} coordinates with the mutual information between the BAT distance and either BAT angle. Figure \ref{fig:MI} shows that, with few exceptions at large bin sizes where values have not converged, the BAT coordinates exhibit lower mutual information, which demonstrates an increased amount of decoupling. Given these insights on an information-theoretic level, we suggest the presented graphs as strong evidence for the decoupling of the coordinate system as a major factor for the improvements.

\section{Conclusion}
Inspired by recent pioneering work, we have demonstrated a considerable beneficial effect for the application of a well-behaved coordinate system on machine learning-based molecular linker generation.
The enhancements of our proposed relative coordinate system, which roots in the bond-angle-torsion (BAT) coordinate formalism, were established by comparing to said pioneering work. In this context, we roughly halved (a reduction of 44.3 percent) the number of test set examples for which the ground-truth linker could not be recovered. A comparative analysis on various indicative aspects was performed. First, commonly used metrics for generative models as well as molecular graph evaluation were presented. Furthermore, ablation studies on the included coordinate types were compared in order to identify the coordinates which provide the most valuable information to the model. The reasons for the performance of the different coordinates were elaborated by performing a robustness analysis with respect to the conformer generation process given different random seeds. By means of information-theoretical calculations, the amount of decoupling in the two input coordinate sets was compared. The results support the hypothesis that the decoupled nature of our presented coordinate system plays a major role for the improved performance of the generative process. 

The application of structural information for future state-of-the-art models in molecular fragment linking arguably suggests itself. Given the considerable improvements demonstrated, we propose the presented coordinate system as a standard technique for linking molecular fragments.

\section{Acknowledgements}
We thank Anton A. Polyansky for a critical reading of the manuscript.

\section{Appendix}
\subsection{Transformation from BAT to anchored Cartesian coordinates}
In order to demonstrate the completeness of the proposed BAT coordinate system (Equation \ref{eq:bat2}), the back-transformation to anchored Cartesian coordinates \cite{potter_coordinate_2002, chang_calculation_2003} (Equation \ref{eq:int_cart}) is given as follows.
\begin{align} 
L_1^x &= |\vec{b_1}|\nonumber\\ 
L_2^x &= L_1^x + |\vec{b_2}|\cos\left(\alpha_1\right)\nonumber\\ 
L_2^y &= |\vec{b_2}|\sin\left(\alpha_1\right)\nonumber\\ 
\vec{r}_{E_2} &= \vec{r}_{L_2} + \vec{b_3}'\cos\left(\phi\right)+\left(\frac{\vec{b_2}}{|\vec{b_2}|}\times\vec{b_3}'\right)\sin\left(\phi\right)\nonumber\\
&+\left(\frac{\vec{b_2}^{\intercal}}{|\vec{b_2}|}\cdot\vec{b_3}'\right)[1-\cos\left(\phi\right)]\frac{\vec{b_2}}{|\vec{b_2}|}\nonumber
&\nonumber\\
\text{with}&\nonumber\\
\vec{b_3}'^{\intercal} &=\left(|\vec{b_3}|\cos\left(\alpha_1+\alpha_2\right),\;\; |\vec{b_3}|\cos\left(\alpha_1+\alpha_2\right),\;\;0\right)\nonumber\\
\label{eq:backtrafo}
\end{align}
Here, $\vec{b_3}'$ represents $\vec{b_3}$ for the case $\phi=0$. In order to obtain $\vec{b_3}$ from $\vec{b_3}'$, we rotate $\vec{b_3}'$ around an axis parallel to $\vec{b_2}$ by an angle $\phi$. Following previous work \cite{parsons_practical_2005}, this coordinate transformation is applied by using Rodrigues' rotation formula. Note that Equation \ref{eq:backtrafo} is never explicitly calculated. It is given here for the sole purpose of demonstrating the equivalence of the BAT and internal Cartesian coordinate systems.

\bibliographystyle{unsrt}
\bibliography{references}

\end{document}